\documentclass[a4paper, 11pt]{report}

\usepackage[dvips]{graphicx}
\usepackage{hyperref}
\usepackage{array}
\usepackage{booktabs}
\usepackage{rotating}
\usepackage{amsmath,amssymb}
\usepackage{fancyhdr}

\newcolumntype{P}[2]{%
  >{\begin{turn}{#1}\begin{minipage}{#2}\small\raggedright\hspace{0pt}}l%
  <{\end{minipage}\end{turn}}%
}

\newcommand{\exce}{$\circledcirc$}
\newcommand{\good}{$\circ$}
\newcommand{\aver}{$\vartriangle$}
\newcommand{\poor}{$\times$}

\begin{document}
\title{Requirement Analyses and Evaluations of Blockchain Platforms per
Possible Use Cases}
\author{Kenji Saito\footnote{Graduate School of Business and Finance,
Waseda University},
Akimitsu Shiseki\footnote{Technical Consultant},
Mitsuyasu Takada\footnote{IBM Japan Ltd.},\\
Hiroki Yamamoto\footnote{CUBE SYSTEM Inc.},
Masaaki Saitoh\footnotemark[4],
Hiroaki Ohkawa\footnotemark[4],
Hirofumi Andou\footnote{IT-One Co.,Ltd.},\\
Naotake Miyamoto\footnote{Tokyo Technical Consultant Co.,Ltd.},
Kazuaki Yamakawa\footnotemark[6],
Kiyoshi Kurakawa\footnotemark[6],\\
Tomohiro Yabushita\footnote{CAC Corpotation},$\:$
Yuji Yamada\footnotemark[7],$\:$\\
Go Masuda\footnote{BlockchainHub Inc.},$\:$
and Kazuyuki Masuda\footnotemark[8]
}
\date{
\vspace{12pt}
\begin{tabular}{lrl}
\multicolumn{1}{l}{Date}&
\multicolumn{1}{l}{Version}&
\multicolumn{1}{l}{Change}\\\hline
2021-03-05&		1.0.0&	Initial version.\\
\hline
\end{tabular}
}

\maketitle

\begin{abstract}
It is said that blockchain will contribute to the digital transformation of
society in a wide range of ways, from the management of public and private
documents to the traceability in various industries, as well as digital
currencies.
A number of so-called blockchain platforms have been developed, and
experiments and applications have been carried out on them.
But are these platforms really conducive to practical use of the blockchain
concept?

To answer the question, we need to better understand what the technology
called blockchain really is.
We need to sort out the confusion we see in understanding what blockchain was
invented for and what it means.
We also need to clarify the structure of its applications.

This document provides a generic model of understanding blockchain and its
applications.
We introduce 4 design patterns to classify the platforms:
blockchain, state machine replication (SMR),
blockchain-centric multi-ledgers (BCML) and decentralized multi-ledgers (DML).
We categorize possible use cases by identifying the structure among
provenance, token and smart contract applications, and organize the functional,
performance, operational and legal requirements for each such case.

Based on the categorization and criteria, we evaluated and compared the
following platforms:
Hyperledger Fabric, Hyperledger Iroha, Hyperledger Indy, Ethereum,
Quorum/Hyperledger Besu, Ethereum 2.0, Polkadot, Corda and BBc-1.
We have tried to be fair in our evaluations and comparisons, but we also
expect to provoke discussion.

The intended readers for this document is anyone involved in development of
application systems who wants to understand blockchain and their platforms,
including non-engineers and non-technologists.
The assessments in this document will allow readers to understand the
technological requirements for the blockchain platforms, to question existing
technologies, and to choose the appropriate platforms for the applications
they envision.
The comparisons hopefully will also be useful as a guide for designing new
technologies.

\begin{description}
\item[Keywords:] 
Blockchain, Ledger, DLT, Dapps, Provenance, Tokens,\\
Smart Contracts, Types of Blockchain, Applications of Blockchain
\end{description}
\end{abstract}

\pagenumbering{roman}

\addcontentsline{toc}{chapter}{Table of Contents}
\tableofcontents

\pagebreak

\addcontentsline{toc}{chapter}{List of Figures}
\listoffigures

\pagebreak

\addcontentsline{toc}{chapter}{List of Tables}
\listoftables

\pagebreak

\pagenumbering{arabic}

\chapter{Introduction}\label{chap-intro}
\section{Motivation}
Even though its reputation as a technology has yet to be established,
it is often said that blockchain will contribute to the digital transformation
of society in a wide range of ways, from the management of public and private
documents to the traceability in various industries, as well as digital
currencies.
A number of so-called blockchain platforms have been developed, and
applications and experiments have been carried out on them.
But are these platforms really conducive to practical use of the blockchain
concept?

To answer the question, we need to better understand what the technology
called blockchain really is.
We need to sort out the confusion we see in understanding what blockchain was
invented for and what it means.
We also need to clarify the structure of its applications.

The origin of the blockchain is the invention of
Bitcoin\cite{Nakamoto2008:Bitcoin}, which is thought to
have been created not to let anyone stop you from transferring your
financial assets as you see fit.
The Bitcoin blockchain, which was invented with the aim of fulfilling that
goal, was supposedly designed to make it provable to all participants that
a digitally signed record of a transaction is unshakably positioned in a
particular past, and therefore the fact of the monetary transfer cannot be
reversed in any way.
If the application, whatever it is, does not require such a proof, it would be
pointless to apply this technology.
Perhaps we need to go back to this point of origin, in order to make a fair
judgment of the available technologies and platforms.

With this in mind, we have conducted requirement analyses and evaluations of
so-called blockchain platforms per possible use cases that require proofs.
The evaluated platforms are
Hyperledger Fabric\cite{Hyperledger:Fabric},
Hyperledger Iroha\cite{Hyperledger:Iroha},
Hyperledger Indy\cite{Hyperledger:Indy},
Ethereum\cite{Buterin2013:Ethereum}\cite{Ethereum:Go},
Quorum\cite{ConsenSys:Quorum}/Hyperledger Besu\cite{Hyperledger:Besu},
Ethereum 2.0\cite{Ethereum:eth20Spec},
Polkadot\cite{Wood2016:Polkadot}\cite{ParityTechnologies:Polkadot},
Corda\cite{Hearn2019:Corda}\cite{Corda:Corda} and
BBc-1\cite{Saito2017:BBc1}\cite{BBc1:BBc1}.

\section{Contributions}
The contributions of this work are as follows:
\begin{enumerate}
\item We categorized the existing platforms into 4 design patterns.
\item We classified applications of blockchain and its possible use cases.
\item We set functional, performance, administrative and compliance criteria
to be satisfied for each use case category.
\item We evaluated the existing platforms according to the criteria, based on
the design patterns and how the technologies of these platforms are governed.
\end{enumerate}

Since our analyses were performed at a high level of abstraction, we did not
see significant differences among use cases in our evaluation, but it should
be useful to see the overall trend.

\section{Document Organization}
The rest of this document is organized as follows:

Chapter~\ref{chap-background} gives background knowledge of how blockchain
technology can be understood.
Chapter~\ref{chap-platforms} introduces platforms we evaluated and compared in
this work.
Chapter~\ref{chap-classify} proposes classification of blockchain applications.
Chapter~\ref{chap-criteria} proposes criteria for each use case category.
Chapter~\ref{chap-comparison} evaluates and compares the platforms.
Chapter~\ref{chap-related} compares this work with other attempts to classify
and/or evaluate blockchain platforms.
Finally, chapter~\ref{chap-outro} gives conclusive remarks.


\chapter{Background}\label{chap-background}

For more formal discussion of how blockchain technology can be understood,
please refer to \cite{Saito2021:LastWillTest} (which is in preparation as of
March 2021).

\section{Blockchain Properties}
As described before, blockchain was first invented in order to realize
Bitcoin, whose goal can be described as follows:
\begin{description}
\item[The Goal of Bitcoin (censorship prevention and resilience):] $\:$\\
{\em To create an asset transfer system in which nobody can stop you from
transferring your own financial assets (in particular, bitcoins) as you see
fit.}
\end{description}
This also suggests that {\em your own financial assets are protected from
other users}, because the fact that the asset was transferred to you in the
past must be secure.

The Bitcoin blockchain has to implement a state machine (state transition
system) that satisfies the following three properties in order to achieve the
above mentioned goal:

\begin{description}
\item[BP-1 (censorship prevention in authentication):] Only a self-authorized
user alone can cause a state transition that is allowed in the state machine.
\item[BP-2 (censorship prevention and fault tolerance):] Such a state
transition always occurs if the authorized user wants it to happen.
\item[BP-3 (censorship prevention for past records):] Once a state transition
occurs, it is virtually irreversible.
\end{description}
where BP stands for {\em Blockchain Property}.

Essentially speaking, the functionality of a platform that advocates
blockchain must achieve all of these BPs, and it should be designed
accordingly.
However, depending on the needs of the business, some or all of these
properties may be relaxed in reality, which may be obscuring the nature of
this technology and keeping people from understanding it accurately.

\section{Clearing up Common Misconceptions}
The following discourses are common in the press and in casual discussions,
but they are all false:
\begin{enumerate}
\item Blockchain is secure because it is encrypted.
\item Blockchain builds consensus (among people).
\item The primary purpose of blockchain is to share information.
\end{enumerate}
We consider that these misconceptions are harmful for reaching accurate
understanding of the technology, and therefore they need to be cleared up
at this moment.
We also hope that this will help clarifying what this technology is and what
it is not.

\subsection{Misconception 1 : It is Encrypted}
Encryption, in short, is the procedure of transforming data so that only
those who know the key can decrypt it.
When looking at the blockchain ecosystem as a whole, private keys are
generally encrypted because users need to keep their private keys hidden so
that only they can use them, as described below.
However, the blockchain itself, which is the mechanism for maintaining
records, is generally not encrypted.
This is because, in order to satisfy BP-\{1, 2, 3\} above, it must be
verifiable by everyone, not just by a specific party who knows the key, that
the record is maintained correctly.
If only a specific party can verify it, then if that party denies it, the
state transition (such as asset transfer) will be stopped\footnote{
Some blockchain platforms allow users to use zero-knowledge proofs to hide
the contents of transactions while ensuring their verifiability by all
(cf. zk-SNARKs\cite{10.1145/2090236.2090263} and
zk-STARKs\cite{DBLP:journals/iacr/Ben-SassonBHR18}).}.

Blockchain makes heavy use of cryptographic hash functions, which are not
encryption because their outputs (digests) cannot be decrypted.
Anyone can calculate the digests, and check if they match the safely stored
values, so that anyone can participate in the verification process.

Many people seem to think that blockchain can protect their privacy, but that
is not the purpose of blockchain, and needs to be worked out to make it a
reality (cf. mixing services, which should be used with
care\cite{10.1145/2896384}).

\subsection{Misconception 2 : It Builds Consensus (among people)}
The term {\em consensus}, which is frequently used to describe blockchain
technology, does not describe an agreement between people.
It is a computer science term that describes the automatic matching of states
(set of variables with values) in multiple processes (and thus does not go so
far as to ``build a consensus'' in the human sense).

The reason why we need consensus in blockchain is that we want to replicate
the state machine onto multiple computers.
The states of the replicated state machines must always match
(hence the consensus).
There are two reasons to replicate a state machine.
One is to achieve fault tolerance in order to satisfy BP-2, i.e., to make
copies of the state machine so that it can continue to function even if some
of the computers stop.
The other is to ensure that the correct records are maintained by comparing
replicas with one another to satisfy BP-3, which, we believe, needs to be
understood carefully.

In many casual discussions, it is understood that {\em everyone sees the same
thing, so it cannot be tampered with (tampering is easily detected)}, but
this is wrong.
It leaves out the important point of what correct records are and which is
correct when there are different (contradicting) records.
Simply deciding by majority vote is not acceptable, because it would allow
someone having many copies (which is relatively inexpensive) to control
correctness.

In fact, blockchain works by the rule that the history that is the most
difficult to falsify is the most correct.
In other words, we can presume that we are looking at the same thing
because it is difficult to falsify.
So many casual discussions are backwards in their understanding.

\subsection{Misconception 3 : Its Primary Purpose is to Share}
If the state machine is replicated with tamper resistance, we can say that
information is shared because the state is shared, and we can assume that we
have a correct copy because it is difficult to tamper with.
This may be useful for many applications.
However, this is not the purpose of blockchain, but a method it uses.

If the information is not shared and is monopolized, then the records could be
controlled by the monopolizing party.
There is always possibility of censorship, and if you cannot detect what is
going on inside in the first place, it does not matter if the state machine is
advertised to be tamper-resistant or not.
The purpose of the blockchain is to make it impossible for any particular
party to control the state machine, by having everyone participate in
verifying the operations of the machine.

\section{How to Design Blockchain}
\begin{figure}[h!]
\begin{center}
\includegraphics[scale=0.48]{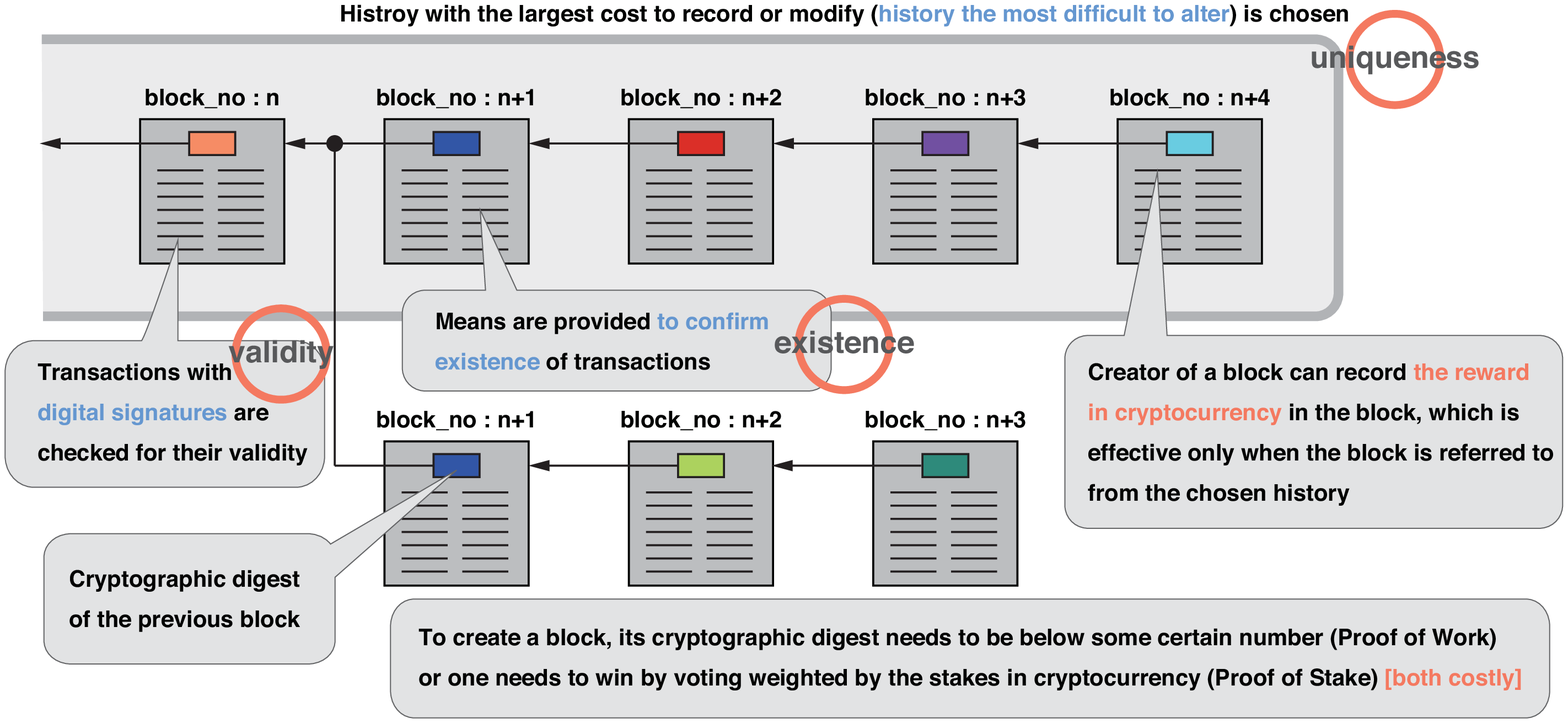}\\
\caption{Abstracted Blockchain}\label{fig-abstracted-blockchain}
\end{center}
\end{figure}

So how can we design a blockchain that satisfies BP-\{1, 2, 3\}?
Figure~\ref{fig-abstracted-blockchain} shows an abstract example.

\subsection{Validity and Availability}
Let us start with BP-1 (censorship prevention in authentication).
When we use our bank cash card to make a transfer, we punch in our PIN at the
ATM.
But then it is the bank that confirms our identity, and the mechanism allows
the bank to stop the transaction at will.
In blockchain, instead, a person's identity is proven by digitally signing a
transaction using a private key.
An unspecified number of people need to verify the digital signature, but
asset transfers and other operations can be performed without the need for a
specific third party.

In order to realize this, however, we need a digital signature mechanism that
can prove the public key used for verification is legitimate without relying
on a public key certificate, which is usually issued by a certificate
authority.
It is required because we need to satisfy BP-1 (suggesting a need for proving
the party's identity without relying on a specific third party) and BP-2
(suggesting a need for proving the party's identity regardless of the validity
period of the certificate).

A typical solution is that a cryptographic digest of a public key is used as
the identifier of an account.
This solution is groundbreaking in that it embeds the information necessary to
verify the signature (the public key) in the data structure.
A public key is attached to the transaction data along with a digital
signature, and the public key is considered legitimate if the digest
calculated from the public key is equal to the address of the authorised user's
account.
This makes it possible for a completely unrelated third party to verify the
legitimacy of the public key, and verify the formal authenticity of the
transaction.

For fault-tolerance part of BP-2, as mentioned earlier, the system is made
redundant by replicating the state machine, allowing the state machine to
operate until the last processor stops\footnote{Of course, if the number of
participants is small, it can be said that the purpose of blockchain will not
be achieved, because it would be easy for a small number of participants to
control the state machine.}.

\subsection{(Difficulty in Disproving) Existence}
BP-3 (censorship prevention for past records) concerns the difficulty of
tampering.
In the case of Bitcoin, the data of transactions is stored in chunks called
blocks, which are created by many participants competing with one another,
and are created on average every ten minutes.
To create a block, you have to win a special lottery (configuring the data
for the block so that its cryptographic digest is smaller than a certain
value, which only succeeds on a completely probabilistic basis).
The more lots you draw, the higher the chance of winning, but the more power
this draw takes, the more electricity you need to pay for.
This is called {\em proof of work}, and the evidence of winning the lottery
goes into the block.
More specifically, the cryptographic digest of the created block is stored in
the next block that follows.

In other words, blocks are formed by referring to their previous blocks,
and are like strung together in a string of beads (or a chain).
If we try to change the record stored in a past block, the evidence of
winning the lottery will be invalid for all subsequent blocks (since their
digests will change).
Thus, the lottery for each block would have to be redrawn, which would
require huge power costs and computational resources.
Since anyone can verify that the blockchain is made of only the blocks that
won the lottery, tampering is said to be virtually impossible.

This mechanism allows us to think that what is being recorded is real.

\subsection{Uniqueness}
Since the creation of blocks is autonomous and competitive, it is possible
for a chain of blocks to split into several chains at some point.
In that case, the chain that is the most difficult to tamper with is
considered to be the only correct chain.
Since tampering with the chain requires redrawing the same lottery, the chain
with the largest accumulated cost of proof of work, i.e., the history with the
most (presumably) drawn lots, is adopted.

\subsection{Source of Protection}
Because of the huge power costs of lotteries when competition increases, some
recent blockchains have adopted {\em proof of stake} instead of lotteries
(proof of work), in which the correct history is determined by a vote weighted
by deposits of native currency tokens.

In the case of proof of work, the cost of electricity for the lottery (the
higher the cost, the more difficult it is to tamper with) is balanced by the
expected market value of the native currency.
Likewise in the case of proof of stake, since the cost of native tokens
that can be deposited (the more, the harder it is to tamper with) depends on
the market value of the currency in question, it can be said that both these
mechanisms depend on the market value of the native currency anyway, and are
protected only while its price is high.

\section{Design Patterns of Blockchain}
\begin{figure}[h!]
\begin{center}
\includegraphics[scale=0.62]{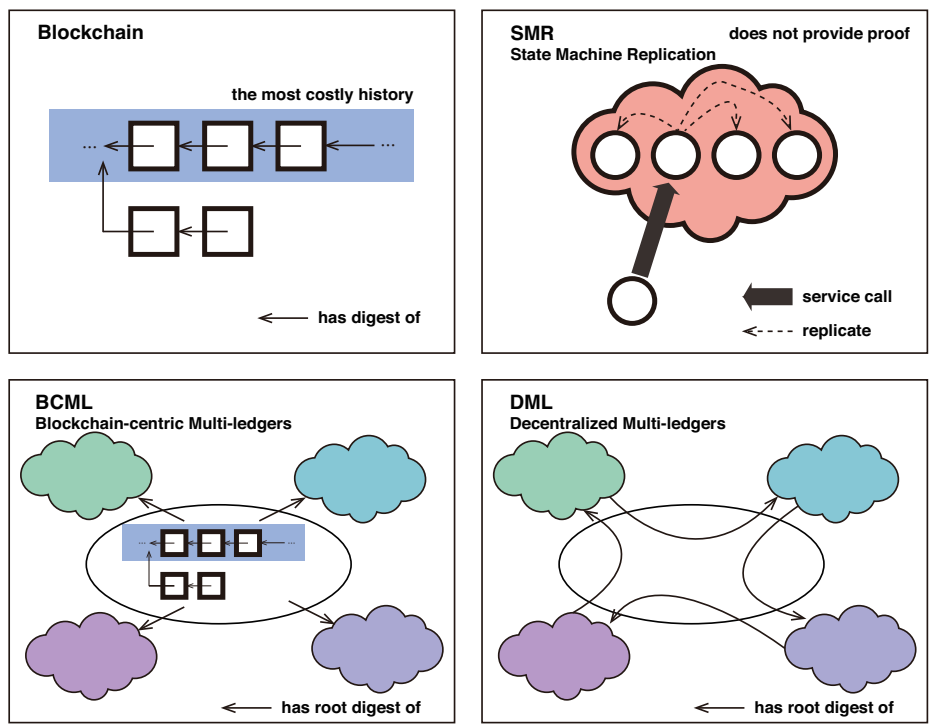}\\
\caption{Design Patterns of Blockchain}\label{fig-design-patterns}
\end{center}
\end{figure}

Platforms that advocates blockchain today can be categorized into four design
patterns as illustrated in Figure \ref{fig-design-patterns}.

\subsection{Blockchain}
Blockchain is as we have described it, and includes many of so-called
{\em public chains} or {\em public ledgers}, which are {\em permissionless} to
join.
It achieves censorship prevention and resilience by allowing participants to
move in and out freely, by packing self-sufficient, verifiable transactions
into blocks, and by assuming that the chain of blocks with the highest
creation cost is the correct history, as it is the most difficult to tamper
with.

\subsection{State Machine Replication (SMR)}
SMR includes many of so-called {\em private chains}, {\em consortium chains} or
{\em private ledgers}, which are {\em permissioned} to join.
In SMR, the participants share a history of events by having replicas of a
single state machine within a controlled membership.
It notably raises two questions:
\begin{enumerate}
\item Do all stakeholders have replicas (or part of it) for verification?

Typically, as Figure \ref{fig-design-patterns} shows,
clients do not have replicas.
Therefore they cannot verify whether the records in the service are correct or
not.
The clients have no choice but to trust the service, and the situation is no
different from existing services that do not use blockchain.

\item Is there a way to verify that the replica one party has is genuine?

Even among members who have replicas, there is no way to be certain that the
state machines they have match and all records are correct.
This is because you cannot exclude the possibility that a different replica
may have been sent only to you.

\end{enumerate}
Because of these problems, SMR cannot really work as blockchain that satisfies
BP-1 and BP-2 especially because it is permissioned (censorship is not
prevented), nor it satisfies BP-3 without an external point of trust.

\subsection{Blockchain-centric Multi-ledgers (BCML)}
BCML is a set of ledgers backed by blockchain.
Each ledger typically forms a growing Merkle tree, and periodically registers
its root to blockchain (the technique often called {\em anchoring}).

A single private ledger anchoring onto a single blockchain is a special case
of this pattern.

\subsection{Decentralized Multi-ledgers (DML)}
DML is like BCML, only that there is no blockchain in the middle.
Anchoring by ledgers is redundantly performed against one another.

Because it does not use blockchain, the source of protection for this
mechanism is the difficulty of recursively breaking into different domains,
and it would be more free from externalities, unlike the strength of
blockchain, which depends on the market price of their native currency.

\section{Proof Problems and the Last Will Test}
Blockchain is often used in combination with digital signatures.
We believe that the real value of blockchain is to solve the essential
challenges of digital signatures.
The challenges are twofold.
One is the inability to verify the correctness of past digital signatures
({\em elapsed-time proof problem}), given the risk of private key
compromise, expiration of the public key certificate, or compromise of the
signature algorithm.
The other is the inability to prove the absence of signatures that were not
really present in the past ({\em alibi proof problem}) even when we could
assume a presence of a ledger where all interested signatures are recoreded,
for the same reasons.

Solving these problems has been clearly important for Bitcoin, for example:
the reason why the initial series of bitcoin transfers when the system went
live in 2009 is still valid 12 years later is that the digital signatures on
the historical transaction data can still be verified as correct today.
In addition, it is not possible to fabricate a transaction that did not
actually occur in 2009.

The first author of this paper is proposing {\em the last will
test} \cite{Saito2021:LastWillTest} as a way to see if the technology
advocating blockchain meets this true value.
A last will should be able to be created and updated if the person wants to,
and no one should be able to stop them from doing so (censorship prevention and
resilience as the goals of blockchain).
Even if the will as a digital document is digitally signed with the person's
private key, it is only after the death of the person that the signature needs
to be verified to make sure it is authentic (elapsed-time proof problem).
It is conceivable that some malicious heir may find the private key in, say,
a USB memory of the deceased, and rewrite the will and digitally sign it
again.
In such a case, we should be able to disprove the autenticity of the document
(alibi proof problem).

For a platform that advocates blockchain, these must be possible without a
single doubt.
For example, if the cryptographic digest of the document and the digital
signature were both embedded in Bitcoin, in a past block before the person's
death, it would then become provable that the document existed as is before
the death of the deceased, not to mention that no one could stop the person
casting transactions to embed the digets and signatures at will.


\chapter{Blockchain Platforms}\label{chap-platforms}

\section{Platforms to be Compared}
Table \ref{tab-compared-platforms} shows the list of blockchain platforms
compared in this work.

\begin{table}[!h]
\begin{center}
\caption{Compared Platforms}
\label{tab-compared-platforms}
{\small
\begin{tabular}{l|l|l|l}\hline
\multicolumn{1}{c|}{Name}&
\multicolumn{1}{c|}{Pattern}&
\multicolumn{1}{c|}{Source of Protection}&
\multicolumn{1}{c}{Community}\\\hline

Hyperledger Fabric&SMR&published state\footnotemark&large\\\hline
Hyperledger Iroha&SMR&membership&moderate\\\hline
Hyperledger Indy&SMR&trust anchors&moderate\\\hline
Ethereum (1.0)&Blockchain&price of ETH&large\\\hline
Quorum/Hyperledger Besu&SMR&membership&large\\\hline
Ethereum 2.0&BCML&price of ETH2&moderate\\\hline
Polkadot&BCML&price of DOT&moderate\\\hline
Corda&SMR&notaries&large\\\hline
BBc-1&DML\footnotemark&recursive anchors&small\\\hline

\end{tabular}
}
\end{center}
\end{table}

\footnotetext[1]{It allows consortium members to publish states.}
\footnotetext{It starts as BCML anchored on Ethereum or
Bitcoin as its ledger subsystem\cite{BBc1:LedgerSubsystem}.}

In the end, what we should hope for from these platforms is that they will
sufficiently satisfy BP-\{1, 2, 3\} and thus solve elapsed-time and alibi
proof problems of digital signatures.
To that end, the table covers all design patterns, and list the source of
protection to show where attacks can be made to disable the protection.
The size of the community is also important to see how it develops in the
future.

Those platforms with their community size ``large'' are probably the ones
you see the most in blockchain demonstrations and products today.

\section{Hyperledger Fabric}
Hyperledger\cite{Linux:Hyperledger}
is a project started in 2015 by the Linux Foundation with a fourfold
mission: 1) prepare a business-ready open source distributed ledger framework
and code
base; 2) create a technical community for open source development; 3) involve
the leaders of the
ecosystem including developers, service/solution providers, and customers;
and 4) provide a
platform for governance.

To achieve these missions, Hyperledger has many open source software
development projects underway.

Hyperledger Fabric\cite{Hyperledger:Fabric},
is probably the most commonly known general-purpose private ledger in
Hyperledger project, whose initial code was a merge between code provided from
IBM and Digital Asset Holdings.

Hyperledger Fabric is SMR by the design pattern.
In response to criticism that the private (or consortium) ledger system does
not guarantee verifiability, it has a mechanism that allows consortium members
to publish the state of the state machine to the Web.

\section{Hyperledger Iroha}
Hyperledger Iroha\cite{Hyperledger:Iroha},
is another general-purpose private ledger in Hyperledger project, whose
initial code was provided from Soramitsu, a startup based in Japan.

Hyperledger Iroha is SMR by the design pattern.
It is lightweight, and designed with a variety of environments in mind,
including mobile devices.

\section{Hyperledger Indy}
Hyperledger Indy\cite{Hyperledger:Indy} is a ledger in Hyperledger proejct
that focuses on realization of decentralized and self-sovereign digital
identities.

Hyperledger Indy is also SMR by the design pattern.
The source of protection for its ledger system itself is the membership
management.
The integrity of digital identities is protected in the end by trust anchors
who the ledger had already known.
Since trust anchors are also under the membership of the ledger, the whole
system can be considered protected by the membership.

\section{Ethereum}
Ethereum\cite{Buterin2013:Ethereum}\cite{Ethereum:Go} is a foundation for
general applications by extending the concept of blockchain
(and is blockchain by the design pattern).

Applications on Ethereum are called {\em smart contracts}, which are not
necessarily augmented versions of contracts as we see in our social lives,
but automated digital objects with verifiable state transitions.
In Ethereum, each validator (miner) runs EVM (Ethereum Virtual Machine) on
which contracts (application programs) are executed.
Ether (or ETH), the native currency of Ethereum, is generated upon validation
of a block just as with Bitcoin.
Ether is sometimes called {\em cryptofuel} because it is converted to a unit
called {\em gas} required to execute a virtual CPU cycle on EVM.

\section{Quorum and Hyperledger Besu}
GoQuorum\cite{ConsenSys:Quorum} and Hyperledger Besu\cite{Hyperledger:Besu}
are both Ethereum clients, compatible with public and private networks of
Ethereum.

When those clients are networked based on Quorum protocol, an enterprise
version of Ethereum, although it would look like blockchain, the system is
SMR by the design pattern, protected by membership.

\section{Ethereum 2.0}
Ethereum 2.0\cite{Ethereum:eth20Spec} is a new version of Ethereum under
development to solve the following issues of the probabilistic state
machine\cite{Saito2016:PSM} of blockchain: lack of finality and lack of
scalability.
In order to tackle these problems, Ethereum 2.0 is introducing a voting
mechanism among self-nominated parties (proof of stake) and {\em shards}
(horizontal partitions).
In addition, the shards will be able to host ledgers other than those based on
EVM, which would make Ethereum 2.0 BCML by the design pattern.

\section{Polkadot}
Polkadot\cite{Wood2016:Polkadot}\cite{ParityTechnologies:Polkadot} is a
framework to host heterogeneous multiple ledgers, which can connect to
existing blockchain ledgers such as Bitcoin or Ethereum, and can host new
ledgers called {\em parachains}.
The multiple ledgers can interwork through the central chain of blocks called
{\em relaychain}, whose state machine is managed via a BFT (Byzantine Fault
Tolerance) algorithm, which makes Polkadot BCML by the design pattern.

\section{Corda}
Corda\cite{Hearn2019:Corda}\cite{Corda:Corda} is a ledger by R3 (a consortium
of financial institutions) specifically designed for managing agreements
between financial institutions.
Corda has a clear mission of achieving ``what I see is what you see, and we
both know that, and the audit can confirm that'', which is apparently
different from Bitcoin's ``not to let anyone stop you from transferring your
own funds as you see fit'', and has been designed accordingly.

Corda is SMR by the design pattern.
Its source of protection is trust in notaries.

\section{BBc-1}
BBc-1\cite{Saito2017:BBc1}\cite{BBc1:BBc1} is a lightweight toolkit for
private ledgers, starting as a special case of BCML to support anchoring to
Ethereum or Bitcoin.
But the goal of BBc-1 is to be DML by the design pattern, with capabilities of
anchoring among ledgers one another as more and more applications adopt this
mechanism, where the source of protection is the difficulty of recursively
breaking into different domains.


\chapter{Classification of Possible Use Cases}\label{chap-classify}

\section{Applications of Blockchain}
Applications of blockchain can be broadly categorized as tokens (substitute
currencies, tickets, etc.) and provenance (proof of history).
Smart contracts are generally understood to be programs that describe these
applications.

However, smart contracts are in essence a provenance application in themselves,
allowing program code, its execution log, and data changes (state transitions)
to be written and traced in blockchain.
In addition, in public platforms such as Ethereum, payments to miners are made
for executing contracts, which is also an application of tokens.

\begin{figure}[h]
\begin{center}
\includegraphics[scale=0.44]{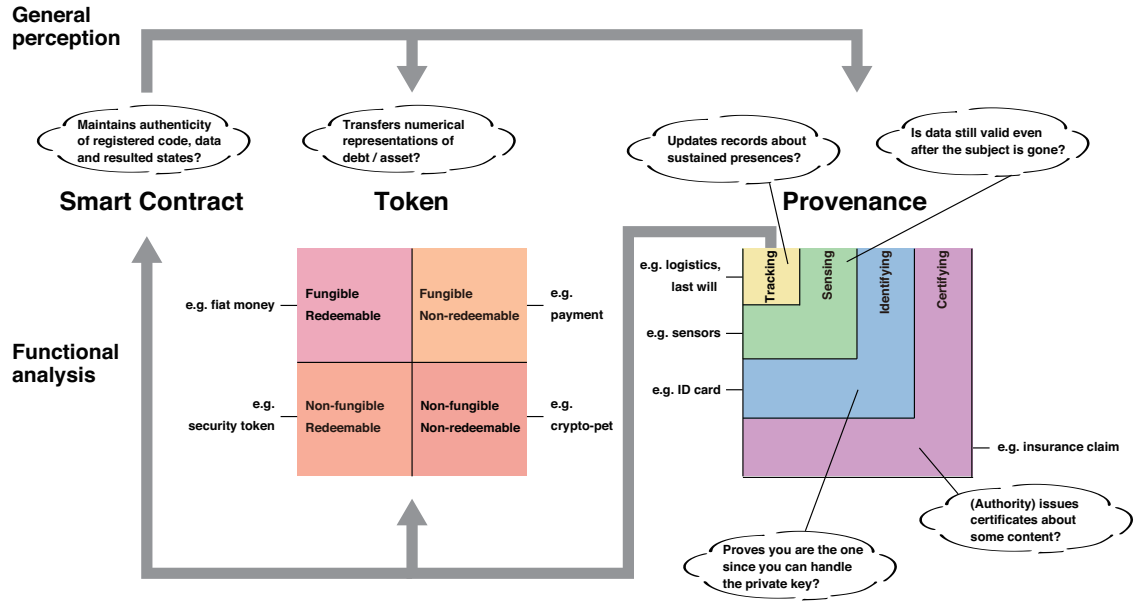}\\
\caption{Classification of Use Cases}\label{fig-classify}
\end{center}
\end{figure}

In this section, we would like to systematically classify the applications of
blockchain, as illustrated in Figure \ref{fig-classify}.

\section{Provenance}

Provenance is to prove the authenticity of some data and its history, and is
represented by four types of applications: certifying, identifying, sensing,
and tracking.

\subsection{Certifying}
A generic model of certifying application is a certificate as a proof of some
content by some certifier.
By writing it to blockchain, the certificate becomes provable of its existence.

Examples include graduation certificates, medical certificates and public key
certificates.

\subsection{Identifying}
When an entity wants to prove its identity, if the certificate for its public
key is on blockchain, it can prove the identity by digitally signing some
message with the corresponding private key.
Therefore, it can be said that identifying is based on certifying
application.

Examples include digital ID cards.

\subsection{Sensing}
Proof of identity can also be applied to sensors, where the sensor itself
holds its private key securely, digitally signs the sensed data, and the data
and signature are written to blockchain, so that the origin of the data can be
permanently proven even if the sensor itself is subsequently destroyed, the
private key is stolen, or the public key certificate expires.

Examples include surveillance cameras.

This can be applied to statements that a person wants to prove that it is
genuinely theirs, such as some political statements, because a person can be
regarded as a sensor that outputs the will of the person.

\subsection{Tracking}
If reliable sensing is possible in this way, tracking will be also made
possible by continuously sensing the state of a persistent entity.

Examples include verifiable GPS tracking of a self-driving cars, and the last
will and testament by a person written and updated at their own will.

\paragraph{}
If we understand the structure of blockchain applications as above, we can
build up all provenance applications based upon the certifying functionality.

\section{Tokens}
Tokens are numerical representations of some liability (which usually means
that the tokens are redeemable) or pure asset (not redeemable), and can be
roughly divided into {\em non-fungible tokens} (irreplaceable; NFT in short)
and {\em fungible tokens} (replaceable)\footnote{Token Taxonomy
Framework\cite{IWA2020:TTF} classifies 5 variables including fungibility and
redeemability.
In this work, we only use the two variables that fundamentally affects the
state machine of tokens}.

\subsection{Fungible or Non-fungible Tokens}
As it is possible to track persistent entities by the tracking functionality,
this directly contributes to the realization of non-fungible tokens.

If we treat a set of non-fungible tokens as a representation of some quantity,
we can realize a fungible token.

Example applications of non-fungibe tokens include tickets and securities.
Those of fungible tokens include payment.

The first application of blockchain, namely Bitcoin, was a system of fungible
tokens, which, as Figure \ref{fig-classify} shows, is at the top of the pile of
features for applications.
The fact that blockchain was designed to implement a fungible token system
like Bitcoin may be what made all of its applications possible, including
provenance and smart contracts.

\subsection{Redeemable or Non-redeemable Tokens}

Redeemability affects the state machine of tokens as redeemable tokens must
disappear upon redemption.

Example applications of redeemable tokens include fiat money and securities.
Those of non-redeemable tokens include cryptocurrencies and crypto-pets.

\section{Smarkt Contracts}
It was mentioned earlier that smart contracts can be seen as an application of
provenance and tokens.
Therefore in the end, as long as there is an application infrastructure for
certifying, all possible applications of blockchain can be covered by
accumulating functions.


\chapter{Criteria}\label{chap-criteria}

\section{Policy for Determining Criteria}
This chapter covers the criteria for evaluating blockchain platforms.
We will focus on practicality as much as possible, and organize the
requirements for functionality, performance, administration and legal
compliance.

The following sections provide a general explanation for each item.
The specific values depend on use cases.

\section{Functional Criteria}

\paragraph{Ability of Proof}
Stakeholders can verify the authenticity of the records without trusting
anyone.

\paragraph{Confidentiality}
Information that shouldn't be known to certain parties is not known to the
parties.

\paragraph{Consistency}
The written record, wherever it is referred to, shows what was written the
last time.

\section{Performance Criteria}

\paragraph{Availability}
Service is not interrupted beyond acceptable limits, i.e., some duration of
time that society can wait.
Some essential applications may require strict 24/7 service.

\paragraph{Response}
Service responds within the allowed time.
Usually for human users, it should respond within 1$\sim$3 seconds.

\paragraph{Throughput}
It handles peak transaction volume.
This depends on the size of the user base as well as other variables, but
for provenance applications, peak at around 1,000tx/s is one standard to think
about, and for token applications, it would be around 100,000tx/s.

\paragraph{Scalability}
It works as we scale up or down the system.
Most applications require dynamic scalability as the user base grows.

\paragraph{Required Resources}
It doesn't consume too much memory/storage, CPU time, or network bandwidth
of components.

\paragraph{Energy Efficiency}
It doesn't consume too much energy.

\section{Administrative Criteria}

\paragraph{Attack Resistance}
It withstands cyber attacks, including DoS (Denial of Service) attacks and
network partitioning.

\paragraph{Ease to Learn}
It is well-documented, and it includes tutorials for self-study.

\paragraph{Ease to Develop}
Libraries are extensive, and the developed code is reusable.

\paragraph{Size of Developer Population}
There is a large population of engineers involved, so that it is easy to obtain
solutions to problems, and easy to hire engineers.

\paragraph{Inter-operability}
It is easy to connect and integrate with other blockchain systems of different
designs.

\paragraph{Ease to Maintain}
It remains backward and upper compatible, and it is easy to update/upgrade.

\paragraph{Ease to Deploy}
It takes only short time from planning to service in.
We can put it on an existing cloud.

\paragraph{Ease to Backup and Restore}
It is easy to make generation-controlled backups and restore from them.

\paragraph{Ease to Automate}
Human resources required for the operation can be made sufficiently small.

\paragraph{Support}
Application developers can get support (for a fee) from the platform developer,
vendor, etc.

\paragraph{Autonomy}
Application planners and developers can avoid vendor lock-in.

\section{Compliance Criterion}

\paragraph{Compliance to Local Regulations}
Service can be made in a law-abiding manner without requiring new legislation
for deployment.

\section{Criteria for Possible Use Cases}

\subsection{How to Think about Possible Use Cases}
There are a wide range of possible use cases, but we will consider typical
use cases for each category.
For provenance applications, one representing use case is picked up for each
category among certifying, identifying, sensing and tracking.
For token applications, two use cases covering two orthogonal axes of
fungibility and redeemability are picked up.

In the table that shows the criteria for each specific use case, only those
items that need special criteria for the case are written.
Blank spaces indicate that general criteria apply.

\pagebreak

\subsection{Certifying}

\paragraph{Insurance Claims}
Service accepts applications for insurance coverage based on a doctor's note
online with their digital signature.

Table~\ref{tab-insurance-claims} shows the specific criteria with the use case
in mind, which should be applicable to other certifying applications as well.

\begin{table}[!h]
\begin{center}
\caption{Specific Criteria : Insurance Claims (Certifying)}
\label{tab-insurance-claims}
{\small
\begin{tabular}{l|l|p{7cm}}\hline
\multicolumn{1}{c|}{Major Cat.}&
\multicolumn{1}{c|}{Minor Cat.}&
\multicolumn{1}{c}{Features to fit the use case}\\\hline
Functional
&Ability of Proof&
It can check the authenticity of the medical reports online.
	\\\cline{2-3}
&Confidentiality&
There is no leakage of information such as personal diagnosis history.
	\\\cline{2-3}
&Consistency&
If a doctor revokes or updates a medical certificate, it is reflected
everywhere.
	\\\hline

Performance
&Availability&
	\\\cline{2-3}
&Response&
	\\\cline{2-3}
&Throughput&
It may peak at around 1,000tx/s.
	\\\cline{2-3}
&Scalability&
	\\\cline{2-3}
&Resources&			\\\cline{2-3}
&Energy&
	\\\hline

Administrative
&Attack Resistance&	\\\cline{2-3}
&Ease to Learn&		\\\cline{2-3}
&Ease to Develop&
Library for creating certificates is available.
	\\\cline{2-3}
&Developers&		\\\cline{2-3}
&Inter-operability&
It can handle medical certificates issued by doctors in other countries and
systems.
	\\\cline{2-3}
&Maintenance&		\\\cline{2-3}
&Ease to Deploy&
	\\\cline{2-3}
&Backup/Restore&	\\\cline{2-3}
&Automation&		\\\cline{2-3}
&Support&			\\\cline{2-3}
&Autonomy&			\\\hline

Compliance
&Local Regulations&
	\\\hline
\end{tabular}
}
\end{center}
\end{table}

\pagebreak

\subsection{Identifying}

\paragraph{Alcohol Shopper}
Service checks the age of the shopper at the shop automatically with their digital
ID card in a personal device (may require biometrics on the device).

Table~\ref{tab-alcohol-shopper} shows the specific criteria with the use case
in mind, which should be applicable to other identifying applications.

\begin{table}[!h]
\begin{center}
\caption{Specific Criteria : Alcohol Shopper (Identifying)}
\label{tab-alcohol-shopper}
{\small
\begin{tabular}{l|l|p{7cm}}\hline
\multicolumn{1}{c|}{Major Cat.}&
\multicolumn{1}{c|}{Minor Cat.}&
\multicolumn{1}{c}{Features to fit the use case}\\\hline
Functional
&Ability of Proof&
It can check the authenticity of the ID card and the possessor.
	\\\cline{2-3}
&Confidentiality&
There is no leakage of information other than the age or the date of birth.
	\\\cline{2-3}
&Consistency&
If the issuer revokes or updates the ID card, it is reflected everywhere.
	\\\hline

Performance
&Availability&
There can be emergencies depending on types of identification.
	\\\cline{2-3}
&Response&
	\\\cline{2-3}
&Throughput&
It may peak at around 1,000tx/s.
	\\\cline{2-3}
&Scalability&
	\\\cline{2-3}
&Resources&			\\\cline{2-3}
&Energy&
	\\\hline

Administrative
&Attack Resistance&	\\\cline{2-3}
&Ease to Learn&		\\\cline{2-3}
&Ease to Develop&
Library for handling public key certificates is available.
	\\\cline{2-3}
&Developers&		\\\cline{2-3}
&Inter-operability&
Service can handle ID cards issued in other locales.
	\\\cline{2-3}
&Maintenance&		\\\cline{2-3}
&Ease to Deploy&
	\\\cline{2-3}
&Backup/Restore&	\\\cline{2-3}
&Automation&		\\\cline{2-3}
&Support&			\\\cline{2-3}
&Autonomy&			\\\hline

Compliance
&Local Regulations&
Privacy regulations need to be taken care of.
	\\\hline
\end{tabular}
}
\end{center}
\end{table}

\pagebreak

\subsection{Sensing}

\paragraph{Surveillance Camera}
Service makes sure that video image sent from a remote camera is genuinely
from that camera.
Authenticity of the video data can be proven even after the camera's private
key is stolen.

Table~\ref{tab-surveillance-camera} shows the specific criteria with the use
case in mind, which should be applicable to other sensing applications.

\begin{table}[!h]
\begin{center}
\caption{Specific Criteria : Surveillance Camera (Sensing)}
\label{tab-surveillance-camera}
{\small
\begin{tabular}{l|l|p{7cm}}\hline
\multicolumn{1}{c|}{Major Cat.}&
\multicolumn{1}{c|}{Minor Cat.}&
\multicolumn{1}{c}{Features to fit the use case}\\\hline
Functional
&Ability of Proof&
Even if the camera is intruded, and the private key is exposed, data written
before the incident can be proven to have been written by the camera.
	\\\cline{2-3}
&Confidentiality&
The video image from the camera is not disclosed to anyone other than the
legitimate users (and the court) even when its authenticity needs to be proven.
	\\\cline{2-3}
&Consistency&
Even if the camera is replaced and a different key pair is used, the output
will be recognized as a series of videos connected from the past.
	\\\hline

Performance
&Availability&
	\\\cline{2-3}
&Response&
	\\\cline{2-3}
&Throughput&
It may peak at around 1,000tx/s.
	\\\cline{2-3}
&Scalability&
	\\\cline{2-3}
&Resources&			\\\cline{2-3}
&Energy&
	\\\hline

Administrative
&Attack Resistance&	\\\cline{2-3}
&Ease to Learn&		\\\cline{2-3}
&Ease to Develop&
It provides an extensive library for creating certificates.
	\\\cline{2-3}
&Developers&		\\\cline{2-3}
&Inter-operability&
It produces data that can be handled in other countries and systems.
	\\\cline{2-3}
&Maintenance&		\\\cline{2-3}
&Ease to Deploy&
	\\\cline{2-3}
&Backup/Restore&	\\\cline{2-3}
&Automation&		\\\cline{2-3}
&Support&			\\\cline{2-3}
&Autonomy&			\\\hline

Compliance
&Local Regulations&
	\\\hline
\end{tabular}
}
\end{center}
\end{table}

\pagebreak

\subsection{Tracking}

\paragraph{Healthcare resource distribution}
Based on the premise of a universal health care system, the insurance card
number and the amount distributed to the individual in the past are linked,
and can be checked at the distribution location (e.g. the case of face mask
distribution in Taiwan or COVID-19 vaccination in many countries).

Table~\ref{tab-healthcare-resource} shows the specific criteria with the use
case in mind, which should be applicable to other tracking applications.

\begin{table}[!h]
\begin{center}
\caption{Specific Criteria : Healthcare Resource Distribution (Tracking)}
\label{tab-healthcare-resource}
{\small
\begin{tabular}{l|l|p{7cm}}\hline
\multicolumn{1}{c|}{Major Cat.}&
\multicolumn{1}{c|}{Minor Cat.}&
\multicolumn{1}{c}{Features to fit the use case}\\\hline
Functional
&Ability of Proof&
Citizens and civil society organizations can verify that the government's
(commissioned) programs are being operated correctly.
	\\\cline{2-3}
&Confidentiality&
Individual's personal medical history will not be exposed.
	\\\cline{2-3}
&Consistency&
If one gets a distribution and then move to a neighboring distribution
location to try, it will return a consistent result.
	\\\hline

Performance
&Availability&
	\\\cline{2-3}
&Response&
	\\\cline{2-3}
&Throughput&
It may peak at around 1,000tx/s.
	\\\cline{2-3}
&Scalability&
	\\\cline{2-3}
&Resources&			\\\cline{2-3}
&Energy&
	\\\hline

Administrative
&Attack Resistance&	\\\cline{2-3}
&Ease to Learn&		\\\cline{2-3}
&Ease to Develop&	\\\cline{2-3}
&Developers&		\\\cline{2-3}
&Inter-operability&
For example, it can be linked to the country's national statistical office
system on other platforms.
	\\\cline{2-3}
&Maintenance&		\\\cline{2-3}
&Ease to Deploy&
It can start serving at the period between the outbreak and the epidemic.
	\\\cline{2-3}
&Backup/Restore&	\\\cline{2-3}
&Automation&		\\\cline{2-3}
&Support&			\\\cline{2-3}
&Autonomy&			\\\hline

Compliance
&Local Regulations&	\\\hline
\end{tabular}
}
\end{center}
\end{table}

\pagebreak

\subsection{Non-fungible, Redeemable Tokens}

\paragraph{Security Tokens}
For example, land is pledged as collateral, and the units that divide the
rights are expressed as tokens, which are put into circulation.

Table~\ref{tab-security-tokens} shows the specific criteria with the use
case in mind, which should be applicable to other redeemable NFT applications.

\begin{table}[!h]
\begin{center}
\caption{Specific Criteria : Security Tokens (Redeemable NFT)}
\label{tab-security-tokens}
{\small
\begin{tabular}{l|l|p{7cm}}\hline
\multicolumn{1}{c|}{Major Cat.}&
\multicolumn{1}{c|}{Minor Cat.}&
\multicolumn{1}{c}{Features to fit the use case}\\\hline
Functional
&Ability of Proof&
It provides the same or greater level of proof than the national land
registry.
	\\\cline{2-3}
&Confidentiality&
No information that should be known only to the involved parties will be
leaked.
	\\\cline{2-3}
&Consistency&
Tokens cannot be double-spend at any point in time (and the transaction is
final).
	\\\hline

Performance
&Availability&
	\\\cline{2-3}
&Response&
	\\\cline{2-3}
&Throughput&
It may peak at around 100,000tx/s.
	\\\cline{2-3}
&Scalability&
	\\\cline{2-3}
&Resources&			\\\cline{2-3}
&Energy&
	\\\hline

Administrative
&Attack Resistance&	\\\cline{2-3}
&Ease to Learn&		\\\cline{2-3}
&Ease to Develop&
An extensive token library is available.
	\\\cline{2-3}
&Developers&		\\\cline{2-3}
&Inter-operability&
Service can have an atomic swap or DVP (Delivery Versus Payment) with other
token systems.
	\\\cline{2-3}
&Maintenance&		\\\cline{2-3}
&Ease to Deploy&
	\\\cline{2-3}
&Backup/Restore&	\\\cline{2-3}
&Automation&		\\\cline{2-3}
&Support&			\\\cline{2-3}
&Autonomy&			\\\hline

Compliance
&Local Regulations&
Transaction can be voided by a court order or other legal reasons.
	\\\hline
\end{tabular}
}
\end{center}
\end{table}

\pagebreak

\subsection{Fungible, Non-redeemable Tokens}

\paragraph{Payment}
Simple sending and receiving coins.

Table~\ref{tab-payment} shows the specific criteria with the use
case in mind, which should be applicable to other non-redeemable, fungible
token applications.

\begin{table}[!h]
\begin{center}
\caption{Specific Criteria : Payment (Non-redeemable Fungible Tokens)}
\label{tab-payment}
{\small
\begin{tabular}{l|l|p{7cm}}\hline
\multicolumn{1}{c|}{Major Cat.}&
\multicolumn{1}{c|}{Minor Cat.}&
\multicolumn{1}{c}{Features to fit the use case}\\\hline
Functional
&Ability of Proof&
It can verify authenticity of the transmissions online.
	\\\cline{2-3}
&Confidentiality&
Information is only exchanged between the parties and is not leaked to others.
	\\\cline{2-3}
&Consistency&
Interactions between the parties are reflected (in the form of changes of their
balances) to the whole as they are.
	\\\hline

Performance
&Availability&
	\\\cline{2-3}
&Response&
	\\\cline{2-3}
&Throughput&
It may peak at around 100,000tx/s.
	\\\cline{2-3}
&Scalability&
	\\\cline{2-3}
&Resources&			\\\cline{2-3}
&Energy&
	\\\hline

Administrative
&Attack Resistance&	\\\cline{2-3}
&Ease to Learn&		\\\cline{2-3}
&Ease to Develop&
It provides an extensive library for creating and operating digital currencies.
	\\\cline{2-3}
&Developers&		\\\cline{2-3}
&Inter-operability&
Service can work with other blockchain networks to have an atomic swap or DVP
with other tokens.
	\\\cline{2-3}
&Maintenance&		\\\cline{2-3}
&Ease to Deploy&
	\\\cline{2-3}
&Backup/Restore&	\\\cline{2-3}
&Automation&		\\\cline{2-3}
&Support&			\\\cline{2-3}
&Autonomy&			\\\hline

Compliance
&Local Regulations&
Transaction can be voided by a court order or other legal reasons.
	\\\hline
\end{tabular}
}
\end{center}
\end{table}


\chapter{Comparison of Platforms}\label{chap-comparison}

\section{Evaluation Method}
In this chapter, we will evaluate specific blockchain platforms according to
the categories and criteria described in the previous chapters.

Table \ref{tab-comparison-method} shows how each major item will be evaluated.

\begin{table}[!h]
\begin{center}
\caption{Evaluation Method}
\label{tab-comparison-method}
{\small
\begin{tabular}{l|p{8cm}}\hline
\multicolumn{1}{c|}{Major Cat.}&
\multicolumn{1}{c}{How to evaluate}\\\hline
Functional&
Based on the design pattern rather than the current implementation.
\\\hline
Performance&
Based on the design pattern rather than the current implementation.
\\\hline
Administrative&
Based on the policy and management capability of the developing organization,
as well as the design pattern.
\\\hline
Compliance&
Based on the policy and management capability of the developing organization,
as well as the design pattern.
\\\hline
\end{tabular}
}
\end{center}
\end{table}

Functional and performance requirements are evaluated for achievability based
on design patterns, assuming that the software will continue to be developed,
rather than on the current implementation of individual platforms.
With regard to administrative and legal requirements, we pay attention to the
types of organizations in which the development is being carried out, and
evaluate them taking into account their policy and management capabilities.

Because of this high level of abstraction, the evaluation of each platform
does not necessarily show significant differences among use case categories.

\pagebreak

\section{Hyperledger Fabric}

Table~\ref{tab-fabric} shows our evaluation of Hyperledger Fabric.

\begin{table}[!h]
\begin{center}
\caption{Hyperledger Fabric}
\label{tab-fabric}
{\small
\begin{tabular}{l|l|c|c|c|c|c|c|l}\hline
\multicolumn{1}{c|}{Major Cat.}&
\multicolumn{1}{c|}{Minor Cat.}&
\multicolumn{1}{P{90}{1.8cm}|}{Certifying}&
\multicolumn{1}{P{90}{1.8cm}|}{Identifying}&
\multicolumn{1}{P{90}{1.8cm}|}{Sensing}&
\multicolumn{1}{P{90}{1.8cm}|}{Tracking}&
\multicolumn{1}{P{90}{1.8cm}|}{NFT}&
\multicolumn{1}{P{90}{1.8cm}|}{Fungible}&
\multicolumn{1}{c}{Note}\\\hline

Functional
&Ability of Proof&\good&\good&\good&\good&\good&\good&
{\footnotesize By replica\footnotemark}
	\\\cline{2-8}
&Confidentiality&\aver&\aver&\aver&\aver&\aver&\aver&
	\\\cline{2-8}
&Consistency&\good&\good&\good&\good&\good&\good&
	\\\hline

Performance
&Availability&\good&\good&\good&\good&\good&\good&
	\\\cline{2-8}
&Response&\good&\good&\good&\good&\good&\good&
	\\\cline{2-8}
&Throughput&\good&\good&\good&\good&\good&\good&
	\\\cline{2-8}
&Scalability&\good&\good&\good&\good&\good&\good&
	\\\cline{2-8}
&Resources&\good&\good&\good&\good&\good&\good&
	\\\cline{2-8}
&Energy&\good&\good&\good&\good&\good&\good&
	\\\hline

Administrative
&Attack Resistance&\good&\good&\good&\good&\good&\good&
	\\\cline{2-8}
&Ease to Learn&\good&\good&\good&\good&\good&\good&
	\\\cline{2-8}
&Ease to Develop&\good&\good&\good&\good&\aver&\aver&
	\\\cline{2-8}
&Developers&\exce&\exce&\exce&\exce&\good&\good&
	\\\cline{2-8}
&Inter-operability&\good&\good&\good&\good&\good&\good&
	\\\cline{2-8}
&Maintenance&\good&\good&\good&\good&\good&\good&
	\\\cline{2-8}
&Ease to Deploy&\good&\good&\good&\good&\good&\good&
	\\\cline{2-8}
&Backup/Restore&\good&\good&\good&\good&\good&\good&
	\\\cline{2-8}
&Automation&\good&\good&\good&\good&\good&\good&
	\\\cline{2-8}
&Support&\exce&\exce&\exce&\exce&\good&\good&
	\\\cline{2-8}
&Autonomy&\good&\good&\good&\good&\good&\good&
	\\\hline

Compliance
&Local Regulations&\good&\good&\good&\good&\good&\good&
	\\\hline
\end{tabular}
}
{\footnotesize
\begin{itemize}
\item[*]
$\circledcirc$: excellent $\;$
$\circ$: good $\;$
$\vartriangle$ average $\;$
$\times$ poor
\end{itemize}
}
\end{center}
\end{table}

\footnotetext{State DB, BC files can be made viewable from the consortium
members' web pages.}

Hyperledger Fabric is a general-purpose ledger, but it does not natively
support tokens.
However, there are samples and tutorials that include
ERC20\cite{Ethereum:ERC20} compliant tokens,
so there should not be any trouble developing them.

\pagebreak

\section{Hyperledger Iroha}

Table~\ref{tab-iroha} shows our evaluation of Hyperledger Iroha.

\begin{table}[!h]
\begin{center}
\caption{Hyperledger Iroha}
\label{tab-iroha}
{\small
\begin{tabular}{l|l|c|c|c|c|c|c|l}\hline
\multicolumn{1}{c|}{Major Cat.}&
\multicolumn{1}{c|}{Minor Cat.}&
\multicolumn{1}{P{90}{1.8cm}|}{Certifying}&
\multicolumn{1}{P{90}{1.8cm}|}{Identifying}&
\multicolumn{1}{P{90}{1.8cm}|}{Sensing}&
\multicolumn{1}{P{90}{1.8cm}|}{Tracking}&
\multicolumn{1}{P{90}{1.8cm}|}{NFT}&
\multicolumn{1}{P{90}{1.8cm}|}{Fungible}&
\multicolumn{1}{c}{Note}\\\hline

Functional
&Ability of Proof&\aver&\aver&\aver&\aver&\aver&\aver&
{\footnotesize By replica}
	\\\cline{2-8}
&Confidentiality&\aver&\aver&\aver&\aver&\aver&\aver&
	\\\cline{2-8}
&Consistency&\good&\good&\good&\good&\good&\good&
	\\\hline

Performance
&Availability&\good&\good&\good&\good&\good&\good&
	\\\cline{2-8}
&Response&\good&\good&\good&\good&\good&\good&
	\\\cline{2-8}
&Throughput&\good&\good&\good&\good&\good&\good&
	\\\cline{2-8}
&Scalability&\good&\good&\good&\good&\good&\good&
	\\\cline{2-8}
&Resources&\good&\good&\good&\good&\good&\good&
	\\\cline{2-8}
&Energy&\good&\good&\good&\good&\good&\good&
	\\\hline

Administrative
&Attack Resistance&\good&\good&\good&\good&\good&\good&
	\\\cline{2-8}
&Ease to Learn&\good&\good&\good&\good&\good&\good&
		\\\cline{2-8}
&Ease to Develop&\good&\good&\good&\good&\good&\good&
	\\\cline{2-8}
&Developers&\aver&\aver&\aver&\aver&\aver&\aver&
		\\\cline{2-8}
&Inter-operability&\good&\good&\good&\good&\good&\good&
	\\\cline{2-8}
&Maintenance&\good&\good&\good&\good&\good&\good&
		\\\cline{2-8}
&Ease to Deploy&\good&\good&\good&\good&\good&\good&
	\\\cline{2-8}
&Backup/Restore&\good&\good&\good&\good&\good&\good&
	\\\cline{2-8}
&Automation&\good&\good&\good&\good&\good&\good&
		\\\cline{2-8}
&Support&\good&\good&\good&\good&\good&\good&
			\\\cline{2-8}
&Autonomy&\good&\good&\good&\good&\good&\good&
			\\\hline

Compliance
&Local Regulations&\good&\good&\good&\good&\good&\good&
	\\\hline
\end{tabular}
}
{\footnotesize
\begin{itemize}
\item[*]
$\circledcirc$: excellent $\;$
$\circ$: good $\;$
$\vartriangle$ average $\;$
$\times$ poor
\end{itemize}
}
\end{center}
\end{table}

Hyperledger Iroha is also a general purpose ledger, and has a proven track
record in token development\cite{Linux:CaseSoramitsu}.
However, the developer community is not large.

\pagebreak

\section{Hyperledger Indy}

Table~\ref{tab-indy} shows our evaluation of Hyperledger Indy.

\begin{table}[!h]
\begin{center}
\caption{Hyperledger Indy}
\label{tab-indy}
{\small
\begin{tabular}{l|l|c|c|c|c|c|c|l}\hline
\multicolumn{1}{c|}{Major Cat.}&
\multicolumn{1}{c|}{Minor Cat.}&
\multicolumn{1}{P{90}{1.8cm}|}{Certifying}&
\multicolumn{1}{P{90}{1.8cm}|}{Identifying}&
\multicolumn{1}{P{90}{1.8cm}|}{Sensing}&
\multicolumn{1}{P{90}{1.8cm}|}{Tracking}&
\multicolumn{1}{P{90}{1.8cm}|}{NFT}&
\multicolumn{1}{P{90}{1.8cm}|}{Fungible}&
\multicolumn{1}{c}{Note}\\\hline

Functional
&Ability of Proof&\aver&\aver&\aver&\aver&\aver&\aver&
{\footnotesize By trust anchor}
	\\\cline{2-8}
&Confidentiality&\good&\good&\good&\aver&\aver&\good&
	\\\cline{2-8}
&Consistency&\good&\good&\good&\aver&\aver&\good&
	\\\hline

Performance
&Availability&\good&\good&\good&\aver&\aver&\good&
	\\\cline{2-8}
&Response&\good&\good&\good&\aver&\aver&\good&
	\\\cline{2-8}
&Throughput&\good&\good&\good&\aver&\aver&\good&
	\\\cline{2-8}
&Scalability&\good&\good&\good&\aver&\aver&\good&
	\\\cline{2-8}
&Resources&\good&\good&\good&\aver&\aver&\good&
	\\\cline{2-8}
&Energy&\good&\good&\good&\aver&\aver&\good&
	\\\hline

Administrative
&Attack Resistance&\good&\good&\good&\aver&\aver&\good&
	\\\cline{2-8}
&Ease to Learn&\aver&\aver&\aver&\aver&\aver&\aver&
		\\\cline{2-8}
&Ease to Develop&\exce&\exce&\good&\aver&\aver&\good&
	\\\cline{2-8}
&Developers&\aver&\aver&\aver&\aver&\aver&\aver&
		\\\cline{2-8}
&Inter-operability&\good&\good&\good&\aver&\aver&\aver&
	\\\cline{2-8}
&Maintenance&\good&\good&\good&\aver&\aver&\good&
		\\\cline{2-8}
&Ease to Deploy&\good&\good&\good&\aver&\aver&\good&
	\\\cline{2-8}
&Backup/Restore&\good&\good&\good&\aver&\aver&\good&
	\\\cline{2-8}
&Automation&\good&\good&\good&\aver&\aver&\good&
		\\\cline{2-8}
&Support&\good&\good&\good&\aver&\aver&\good&
			\\\cline{2-8}
&Autonomy&\good&\good&\good&\aver&\aver&\good&
			\\\hline

Compliance
&Local Regulations&\good&\good&\good&\aver&\aver&\good&
	\\\hline
\end{tabular}
}
{\footnotesize
\begin{itemize}
\item[*]
$\circledcirc$: excellent $\;$
$\circ$: good $\;$
$\vartriangle$ average $\;$
$\times$ poor
\end{itemize}
}
\end{center}
\end{table}

Hyperledger Indy is a ledger that specializes in decentralized
identifiers\cite{Sabadello:21:DI} and verifiable
credentials\cite{Zundel:19:VCD}.
Therefore, it does not support the use of transferring assets, but it does
have a payment interface.

\pagebreak

\section{Ethereum (1.0)}

Table~\ref{tab-ethereum} shows our evaluation of Ethereum (1.0).

\begin{table}[!h]
\begin{center}
\caption{Ethereum (1.0)}
\label{tab-ethereum}
{\small
\begin{tabular}{l|l|c|c|c|c|c|c|l}\hline
\multicolumn{1}{c|}{Major Cat.}&
\multicolumn{1}{c|}{Minor Cat.}&
\multicolumn{1}{P{90}{1.8cm}|}{Certifying}&
\multicolumn{1}{P{90}{1.8cm}|}{Identifying}&
\multicolumn{1}{P{90}{1.8cm}|}{Sensing}&
\multicolumn{1}{P{90}{1.8cm}|}{Tracking}&
\multicolumn{1}{P{90}{1.8cm}|}{NFT}&
\multicolumn{1}{P{90}{1.8cm}|}{Fungible}&
\multicolumn{1}{c}{Note}\\\hline

Functional
&Ability of Proof&\exce&\exce&\exce&\exce&\exce&\exce&
	\\\cline{2-8}
&Confidentiality&\aver&\aver&\aver&\aver&\aver&\aver&
ZoE
	\\\cline{2-8}
&Consistency&\poor&\poor&\poor&\poor&\poor&\poor&
	\\\hline

Performance
&Availability&\aver&\aver&\aver&\aver&\aver&\aver&
{\footnotesize Market risk}
	\\\cline{2-8}
&Response&\poor&\poor&\poor&\poor&\poor&\poor&
	\\\cline{2-8}
&Throughput&\poor&\poor&\poor&\poor&\poor&\poor&
	\\\cline{2-8}
&Scalability&\poor&\poor&\poor&\poor&\poor&\poor&
	\\\cline{2-8}
&Resources&\good&\good&\good&\good&\good&\good&
{\footnotesize By distribution}
	\\\cline{2-8}
&Energy&\poor&\poor&\poor&\poor&\poor&\poor&
{\footnotesize Proof of work}
	\\\hline

Administrative
&Attack Resistance&\aver&\aver&\aver&\aver&\aver&\aver&
{\footnotesize Weak against DoS}
	\\\cline{2-8}
&Ease to Learn&\exce&\exce&\exce&\exce&\exce&\exce&
		\\\cline{2-8}
&Ease to Develop&\exce&\exce&\exce&\exce&\exce&\exce&
	\\\cline{2-8}
&Developers&\exce&\exce&\exce&\exce&\exce&\exce&
		\\\cline{2-8}
&Inter-operability&\good&\good&\good&\good&\good&\good&
	\\\cline{2-8}
&Maintenance&\poor&\poor&\poor&\poor&\poor&\poor&
{\footnotesize Frequent spec changes}
	\\\cline{2-8}
&Ease to Deploy&\good&\good&\good&\good&\good&\good&
	\\\cline{2-8}
&Backup/Restore&\poor&\poor&\poor&\poor&\poor&\poor&
	\\\cline{2-8}
&Automation&\good&\good&\good&\good&\good&\good&
		\\\cline{2-8}
&Support&\good&\good&\good&\good&\good&\good&
			\\\cline{2-8}
&Autonomy&\good&\good&\good&\good&\good&\good&
			\\\hline

Compliance
&Local Regulations&\aver&\aver&\aver&\aver&\aver&\aver&
	\\\hline
\end{tabular}
}
{\footnotesize
\begin{itemize}
\item[*]
$\circledcirc$: excellent $\;$
$\circ$: good $\;$
$\vartriangle$ average $\;$
$\times$ poor
\end{itemize}
}
\end{center}
\end{table}

Ethereum has a strong proof mechanism, but it is backed by the market value of
Ether, and there is market risk.
Ethereum is a global infrastructure, and it is not certain that it or its
applications (smart contracts) can operate legally in all regions.

\pagebreak

\section{Quorum and Hyperledger Besu}

Table~\ref{tab-besu} shows our evaluation of Quorum and Hyperledger Besu.

\begin{table}[!h]
\begin{center}
\caption{Quorum and Hyperledger Besu}
\label{tab-besu}
{\small
\begin{tabular}{l|l|c|c|c|c|c|c|l}\hline
\multicolumn{1}{c|}{Major Cat.}&
\multicolumn{1}{c|}{Minor Cat.}&
\multicolumn{1}{P{90}{1.8cm}|}{Certifying}&
\multicolumn{1}{P{90}{1.8cm}|}{Identifying}&
\multicolumn{1}{P{90}{1.8cm}|}{Sensing}&
\multicolumn{1}{P{90}{1.8cm}|}{Tracking}&
\multicolumn{1}{P{90}{1.8cm}|}{NFT}&
\multicolumn{1}{P{90}{1.8cm}|}{Fungible}&
\multicolumn{1}{c}{Note}\\\hline

Functional
&Ability of Proof&\aver&\aver&\aver&\aver&\aver&\aver&
{\footnotesize By replica}
	\\\cline{2-8}
&Confidentiality&\good&\good&\good&\good&\good&\good&
	\\\cline{2-8}
&Consistency&\aver&\aver&\aver&\aver&\aver&\aver&
	\\\hline

Performance
&Availability&\good&\good&\good&\good&\good&\good&
	\\\cline{2-8}
&Response&\aver&\aver&\aver&\aver&\aver&\aver&
	\\\cline{2-8}
&Throughput&\aver&\aver&\aver&\aver&\aver&\aver&
	\\\cline{2-8}
&Scalability&\aver&\aver&\aver&\aver&\aver&\aver&
	\\\cline{2-8}
&Resources&\aver&\aver&\aver&\aver&\aver&\aver&
	\\\cline{2-8}
&Energy&\good&\good&\good&\good&\good&\good&
	\\\hline

Administrative
&Attack Resistance&\good&\good&\good&\good&\good&\good&
	\\\cline{2-8}
&Ease to Learn&\exce&\exce&\exce&\exce&\exce&\exce&
		\\\cline{2-8}
&Ease to Develop&\exce&\exce&\exce&\exce&\exce&\exce&
	\\\cline{2-8}
&Developers&\exce&\exce&\exce&\exce&\exce&\exce&
		\\\cline{2-8}
&Inter-operability&\good&\good&\good&\good&\good&\good&
	\\\cline{2-8}
&Maintenance&\aver&\aver&\aver&\aver&\aver&\aver&
		\\\cline{2-8}
&Ease to Deploy&\good&\good&\good&\good&\good&\good&
	\\\cline{2-8}
&Backup/Restore&\good&\good&\good&\good&\good&\good&
	\\\cline{2-8}
&Automation&\good&\good&\good&\good&\good&\good&
		\\\cline{2-8}
&Support&\exce&\exce&\exce&\exce&\exce&\exce&
			\\\cline{2-8}
&Autonomy&\good&\good&\good&\good&\good&\good&
			\\\hline

Compliance
&Local Regulations&\good&\good&\good&\good&\good&\good&
	\\\hline
\end{tabular}
}
{\footnotesize
\begin{itemize}
\item[*]
$\circledcirc$: excellent $\;$
$\circ$: good $\;$
$\vartriangle$ average $\;$
$\times$ poor
\end{itemize}
}
\end{center}
\end{table}

It should be noted that while these private network versions of Ethereum have
the advantage of a large development population, their provability is
compromised.
The mechanism is too complex and redundant to run in a private environment, so
performance is not as good as it could be.

\pagebreak

\section{Ethereum 2.0}

Table~\ref{tab-eth2} shows our evaluation of Ethereum 2.0.

\begin{table}[!h]
\begin{center}
\caption{Ethereum 2.0}
\label{tab-eth2}
{\small
\begin{tabular}{l|l|c|c|c|c|c|c|l}\hline
\multicolumn{1}{c|}{Major Cat.}&
\multicolumn{1}{c|}{Minor Cat.}&
\multicolumn{1}{P{90}{1.8cm}|}{Certifying}&
\multicolumn{1}{P{90}{1.8cm}|}{Identifying}&
\multicolumn{1}{P{90}{1.8cm}|}{Sensing}&
\multicolumn{1}{P{90}{1.8cm}|}{Tracking}&
\multicolumn{1}{P{90}{1.8cm}|}{NFT}&
\multicolumn{1}{P{90}{1.8cm}|}{Fungible}&
\multicolumn{1}{c}{Note}\\\hline

Functional
&Ability of Proof&\exce&\exce&\exce&\exce&\exce&\exce&
	\\\cline{2-8}
&Confidentiality&\good&\good&\good&\good&\good&\good&
	\\\cline{2-8}
&Consistency&\aver&\aver&\aver&\aver&\aver&\aver&
$\approx$1 hour for finalization
	\\\hline

Performance
&Availability&\aver&\aver&\aver&\aver&\aver&\aver&
{\footnotesize Market risk}
	\\\cline{2-8}
&Response&\poor&\poor&\poor&\poor&\poor&\poor&
	\\\cline{2-8}
&Throughput&\good&\good&\good&\good&\good&\good&
{\footnotesize 100,000tx/s in future?}
	\\\cline{2-8}
&Scalability&\good&\good&\good&\good&\good&\good&
	\\\cline{2-8}
&Resources&\good&\good&\good&\good&\good&\good&
	\\\cline{2-8}
&Energy&\good&\good&\good&\good&\good&\good&
	\\\hline

Administrative
&Attack Resistance&\aver&\aver&\aver&\aver&\aver&\aver&
	\\\cline{2-8}
&Ease to Learn&\good&\good&\good&\good&\good&\good&
		\\\cline{2-8}
&Ease to Develop&\good&\good&\good&\good&\good&\good&
	\\\cline{2-8}
&Developers&\good&\good&\good&\good&\good&\good&
		\\\cline{2-8}
&Inter-operability&\good&\good&\good&\good&\good&\good&
	\\\cline{2-8}
&Maintenance&\poor&\poor&\poor&\poor&\poor&\poor&
{\footnotesize Frequent changes?}
		\\\cline{2-8}
&Ease to Deploy&\aver&\aver&\aver&\aver&\aver&\aver&
	\\\cline{2-8}
&Backup/Restorey&\poor&\poor&\poor&\poor&\poor&\poor&
	\\\cline{2-8}
&Automation&\good&\good&\good&\good&\good&\good&
		\\\cline{2-8}
&Support&\good&\good&\good&\good&\good&\good&
			\\\cline{2-8}
&Autonomy&\exce&\exce&\exce&\exce&\exce&\exce&
Freedom at shards
			\\\hline

Compliance
&Local Regulations&\aver&\aver&\aver&\aver&\aver&\aver&
	\\\hline
\end{tabular}
}
{\footnotesize
\begin{itemize}
\item[*]
$\circledcirc$: excellent $\;$
$\circ$: good $\;$
$\vartriangle$ average $\;$
$\times$ poor
\end{itemize}
}
\end{center}
\end{table}

This new version of Ethereum is moving towards a structure where multiple
different shards are proven by a core blockchain, which has the potential to
be designed to balance provability with throughput and scalability.
However, there is a market risk as it is supported by the market value of ETH2.

\pagebreak

\section{Polkadot}

Table~\ref{tab-polkadot} shows our evaluation of Polkadot.

\begin{table}[!h]
\begin{center}
\caption{Polkadot}
\label{tab-polkadot}
{\small
\begin{tabular}{l|l|c|c|c|c|c|c|l}\hline
\multicolumn{1}{c|}{Major Cat.}&
\multicolumn{1}{c|}{Minor Cat.}&
\multicolumn{1}{P{90}{1.8cm}|}{Certifying}&
\multicolumn{1}{P{90}{1.8cm}|}{Identifying}&
\multicolumn{1}{P{90}{1.8cm}|}{Sensing}&
\multicolumn{1}{P{90}{1.8cm}|}{Tracking}&
\multicolumn{1}{P{90}{1.8cm}|}{NFT}&
\multicolumn{1}{P{90}{1.8cm}|}{Fungible}&
\multicolumn{1}{c}{Note}\\\hline

Functional
&Ability of Proof&\exce&\exce&\exce&\exce&\exce&\exce&
	\\\cline{2-8}
&Confidentiality&\good&\good&\good&\good&\good&\good&
	\\\cline{2-8}
&Consistency&\aver&\aver&\aver&\aver&\aver&\aver&
	\\\hline

Performance
&Availability&\aver&\aver&\aver&\aver&\aver&\aver&
{\footnotesize Market risk}
	\\\cline{2-8}
&Response&\aver&\aver&\aver&\aver&\aver&\aver&
	\\\cline{2-8}
&Throughput&\good&\good&\good&\good&\good&\good&
	\\\cline{2-8}
&Scalability&\good&\good&\good&\good&\good&\good&
	\\\cline{2-8}
&Resources&\good&\good&\good&\good&\good&\good&
	\\\cline{2-8}
&Energy&\good&\good&\good&\good&\good&\good&
	\\\hline

Administrative
&Attack Resistance&\aver&\aver&\aver&\aver&\aver&\aver&
	\\\cline{2-8}
&Ease to Learn&\good&\good&\good&\good&\good&\good&
		\\\cline{2-8}
&Ease to Develop&\exce&\exce&\exce&\exce&\exce&\exce&
{\footnotesize w/ framework}
	\\\cline{2-8}
&Developers&\good&\good&\good&\good&\good&\good&
		\\\cline{2-8}
&Inter-operability&\exce&\exce&\exce&\exce&\exce&\exce&
{\footnotesize Bridges}
	\\\cline{2-8}
&Maintenance&\good&\good&\good&\good&\good&\good&
		\\\cline{2-8}
&Ease to Deploy&\good&\good&\good&\good&\good&\good&
{\footnotesize w/ framework, needs DOTs}
	\\\cline{2-8}
&Backup/Restore&\poor&\poor&\poor&\poor&\poor&\poor&
	\\\cline{2-8}
&Automation&\good&\good&\good&\good&\good&\good&
		\\\cline{2-8}
&Support&\good&\good&\good&\good&\good&\good&
			\\\cline{2-8}
&Autonomy&\good&\good&\good&\good&\good&\good&
Freedom at parachains
			\\\hline

Compliance
&Local Regulations&\aver&\aver&\aver&\aver&\aver&\aver&
	\\\hline
\end{tabular}
}
{\footnotesize
\begin{itemize}
\item[*]
$\circledcirc$: excellent $\;$
$\circ$: good $\;$
$\vartriangle$ average $\;$
$\times$ poor
\end{itemize}
}
\end{center}
\end{table}

The structure of multiple different parachains being proven by a core
blockchain and connected by bridges to existing blockchains could result in a
design that balances provability, throughput, scalability, and
interoperability.
It also provides a framework that facilitates the building of applications.
However, it is supported by the market value of DOT, so there is a market risk.

\pagebreak

\section{Corda}

Table~\ref{tab-corda} shows our evaluation of Corda.

\begin{table}[!h]
\begin{center}
\caption{Corda}
\label{tab-corda}
{\small
\begin{tabular}{l|l|c|c|c|c|c|c|l}\hline
\multicolumn{1}{c|}{Major Cat.}&
\multicolumn{1}{c|}{Minor Cat.}&
\multicolumn{1}{P{90}{1.8cm}|}{Certifying}&
\multicolumn{1}{P{90}{1.8cm}|}{Identifying}&
\multicolumn{1}{P{90}{1.8cm}|}{Sensing}&
\multicolumn{1}{P{90}{1.8cm}|}{Tracking}&
\multicolumn{1}{P{90}{1.8cm}|}{NFT}&
\multicolumn{1}{P{90}{1.8cm}|}{Fungible}&
\multicolumn{1}{c}{Note}\\\hline

Functional
&Ability of Proof&\aver&\aver&\aver&\aver&\aver&\aver&
{\footnotesize Though auditable}
	\\\cline{2-8}
&Confidentiality&\exce&\exce&\exce&\exce&\exce&\exce&
	\\\cline{2-8}
&Consistency&\good&\good&\good&\good&\good&\good&
	\\\hline

Performance
&Availability&\good&\good&\good&\good&\good&\good&
	\\\cline{2-8}
&Response&\good&\good&\good&\good&\good&\good&
	\\\cline{2-8}
&Throughput&\aver&\aver&\aver&\aver&\aver&\aver&
	\\\cline{2-8}
&Scalability&\good&\good&\good&\good&\good&\good&
	\\\cline{2-8}
&Resources&\good&\good&\good&\good&\good&\good&
	\\\cline{2-8}
&Energy&\good&\good&\good&\good&\good&\good&
	\\\hline

Administrative
&Attack Resistance&\good&\good&\good&\good&\good&\good&
	\\\cline{2-8}
&Ease to Learn&\exce&\exce&\exce&\exce&\exce&\exce&
		\\\cline{2-8}
&Ease to Develop&\good&\good&\good&\good&\good&\good&
	\\\cline{2-8}
&Developers&\exce&\exce&\exce&\exce&\exce&\exce&
{\footnotesize Java/Kotlin}
		\\\cline{2-8}
&Inter-operability&\good&\good&\good&\good&\good&\good&
	\\\cline{2-8}
&Maintenance&\good&\good&\good&\good&\good&\good&
		\\\cline{2-8}
&Ease to Deploy&\good&\good&\good&\good&\good&\good&
	\\\cline{2-8}
&Backup/Restore&\good&\good&\good&\good&\good&\good&
	\\\cline{2-8}
&Automation&\good&\good&\good&\good&\good&\good&
		\\\cline{2-8}
&Support&\exce&\exce&\exce&\exce&\exce&\exce&
			\\\cline{2-8}
&Autonomy&\aver&\aver&\aver&\aver&\aver&\aver&
			\\\hline

Compliance
&Local Regulations&\exce&\exce&\exce&\exce&\good&\good&
Legality of tokens?
	\\\hline
\end{tabular}
}
{\footnotesize
\begin{itemize}
\item[*]
$\circledcirc$: excellent $\;$
$\circ$: good $\;$
$\vartriangle$ average $\;$
$\times$ poor
\end{itemize}
}
\end{center}
\end{table}

It can meet many needs of financial applications because it is designed to be
applied by financial institutions and has large development population.
It does not have a high proof capability, but the emphasis is on being
auditable.
Legal compliance is a particular area of support, but when designing tokens,
care must be taken to ensure that they are legal in the applied regions.

\pagebreak

\section{BBc-1}

Table~\ref{tab-bbc1} shows our evaluation of BBc-1.

\begin{table}[!h]
\begin{center}
\caption{BBc-1}
\label{tab-bbc1}
{\small
\begin{tabular}{l|l|c|c|c|c|c|c|l}\hline
\multicolumn{1}{c|}{Major Cat.}&
\multicolumn{1}{c|}{Minor Cat.}&
\multicolumn{1}{P{90}{1.8cm}|}{Certifying}&
\multicolumn{1}{P{90}{1.8cm}|}{Identifying}&
\multicolumn{1}{P{90}{1.8cm}|}{Sensing}&
\multicolumn{1}{P{90}{1.8cm}|}{Tracking}&
\multicolumn{1}{P{90}{1.8cm}|}{NFT}&
\multicolumn{1}{P{90}{1.8cm}|}{Fungible}&
\multicolumn{1}{c}{Note}\\\hline

Functional
&Ability of Proof&\good&\good&\good&\good&\good&\good&
{\footnotesize (Recursive) anchors}
	\\\cline{2-8}
&Confidentiality&\good&\good&\good&\good&\good&\good&
	\\\cline{2-8}
&Consistency&\good&\good&\good&\good&\good&\good&
	\\\hline

Performance
&Availability&\good&\good&\good&\good&\good&\good&
	\\\cline{2-8}
&Response&\good&\good&\good&\good&\good&\good&
	\\\cline{2-8}
&Throughput&\good&\good&\good&\good&\good&\good&
	\\\cline{2-8}
&Scalability&\good&\good&\good&\good&\good&\good&
	\\\cline{2-8}
&Resources&\exce&\exce&\exce&\exce&\exce&\exce&
	\\\cline{2-8}
&Energy&\exce&\exce&\exce&\exce&\exce&\exce&
	\\\hline

Administrative
&Attack Resistance&\good&\good&\good&\good&\good&\good&
	\\\cline{2-8}
&Ease to Learn&\aver&\aver&\aver&\aver&\aver&\aver&
		\\\cline{2-8}
&Ease to Develop&\good&\good&\good&\good&\good&\good&
{\footnotesize Python/JS/Go}
		\\\cline{2-8}
&Developers&\aver&\aver&\aver&\aver&\aver&\aver&
		\\\cline{2-8}
&Inter-operability&\good&\good&\good&\good&\good&\good&
	\\\cline{2-8}
&Maintenance&\good&\good&\good&\good&\good&\good&
		\\\cline{2-8}
&Ease to Deploy&\good&\good&\good&\good&\good&\good&
	\\\cline{2-8}
&Backup/Restore&\good&\good&\good&\good&\good&\good&
	\\\cline{2-8}
&Automation&\good&\good&\good&\good&\good&\good&
		\\\cline{2-8}
&Support&\good&\good&\good&\good&\good&\good&
			\\\cline{2-8}
&Autonomy&\good&\good&\good&\good&\good&\good&
			\\\hline

Compliance
&Local Regulations&\good&\good&\good&\good&\good&\good&
	\\\hline
\end{tabular}
}
{\footnotesize
\begin{itemize}
\item[*]
$\circledcirc$: excellent $\;$
$\circ$: good $\;$
$\vartriangle$ average $\;$
$\times$ poor
\end{itemize}
}
\end{center}
\end{table}

It is a lightweight framework for general purpose ledgers, and is intended to
be deployed realistically, initially anchored to Ethereum or Bitcoin for
provability, and mutually anchored in the long term to avoid market risk.
However, the development population is small.

\pagebreak


\chapter{Related Work}\label{chap-related}

\section{Categorization of Designs}

The categorization of blockchain designs is often addressed by apparent
differences such as public, consortium, and private types, as shown in 
\cite{m_niranjanamurthy_2018} as a typical example.
But that does not necessarily tell us much about how proof, which is the
heart of the value that blockchain gives, will be provided.
We believe that the four design patterns presented in this document have
clarified what mechanisms can provide proofs, and have shown the
problems with the current designs and where it should go in the future.

\section{Categorization of Use Cases}

\subsection{Categorization of General Use Cases}

\cite{8494045} enumerates use cases of smart contracts as supply chain,
Internet of Things, healthcare systems, digital right management, insurance,
financial systems and real estate.
Likewise, the initial so-called whitepaper by Hyperledger project enumerated
their use cases as financial assets, corporate actions, supply chains, master
data management, sharing economy and Internet of Things.
The problem is that we do not know if these are complete and exhaustive.
In this document, we have attempted to cover all possible kinds of
applications in terms of which functions are used, based on an analysis of the
stack structure of functions of any blockchain platforms.

\subsection{Categorization of Token Systems}

Token Taxonomy Framework\cite{IWA2020:TTF} classifies tokens with five
variables: token type, token unit, value type, representation type and
template type.
These may help those in the token business to understand particular token
systems, but for the most part they do not affect the state machine of tokens.
In this document, we only focused on fungibility and redeemability, keeping in
mind the differences among state machines of tokens.

\section{Evaluation of Platforms}
A number of studies, e.g. \cite{8038517}, have analyzed the performance
differences between blockchain platforms.
While actual measurements will bring valuable information, structural
differences will provide more universal information.

LEAF (LayerX Enterprise blockchain Analysis Framework)\cite{LayerX:LEAF} by
a Japanese blockchain startup LayerX makes comparisons based on understanding
of structures.
In its basic report, Corda, Hyperledger Fabric, and Quorum are compared.
The analyses by LEAF are based on detailed understanding, which are valuable,
but at the same time, it is uncertain whether the obtained knowledge will
remain valid when the platform itself updates.
In this document, we have tried to derive valid knowledge as long as the
basic designs remain the same by evaluating the platforms based on the
properties obtained from the design patterns.


\chapter{Conclusions}\label{chap-outro}

This document provided a generic model of understanding blockchain and its
applications, categorizes possible use cases, and organizes the functional,
performance, operational and legal requirements for each such case.

Based on the categorization and criteria, we evaluated and compared the
following platforms:
Hyperledger Fabric, Hyperledger Iroha, Hyperledger Indy, Ethereum,
Quorum/Hyperledger Besu, Ethereum 2.0, Polkadot, Corda and BBc-1.
We have tried to be fair in our evaluations and comparisons, but we also
expect to provoke discussion.
This work must go on.

The assessments will allow readers to understand the technological requirements
for the blockchain platforms, to question existing technologies, and to choose
the appropriate platforms for the applications they envision.
The comparisons hopefully will also be useful as a guide for designing new
technologies.


\chapter*{Acknowledgment}
We express our thanks to the corporate members of Platform Subcommittee of
BlockchainHub Inc., for provision of valuable discussion and information based
on their experiences with the platforms.

\bibliographystyle{plain}
\bibliography{platforms-comp}

\end{document}